\documentstyle[psfig,11pt]{article}
\begin{document}
\begin{center}
{\Large
A Reverse Monte Carlo study of H+D Lyman alpha absorption from
QSO spectra}
\end{center}

\medskip\noindent
S. A. Levshakov$^{1,2}$,
W. H. Kegel$^{1}$, and
F. Takahara$^{3}$

\medskip\noindent
{\small
$^{1}$Institut f\"ur Theoretische Physik der Universit\"at Frankfurt
am Main, Postfach 11 19 32,\\ 60054 Frankfurt/Main 11, Germany\\
$^{2}$A. F. Ioffe Physico-Technical Institute, 194021 
St. Petersburg, Russia\\
$^{3}$Department of Physics, Tokyo Metropolitan University,
Hachioji, Tokyo 192-03, Japan}

\bigskip\noindent
{\bf Abstract.} A new method based on a Reverse Monte Carlo [RMC] 
technique and aimed at the inverse problem in the analysis 
of interstellar (intergalactic) 
absorption lines is presented. The line formation process in chaotic
media with a finite correlation length 
$(l > 0)$ of the stochastic velocity
field ({\it mesoturbulence}) is considered. This generalizes the standard
assumption of completely uncorrelated bulk motions $(l \equiv 0)$ in the
{\it microturbulent} approximation 
which is used for the data analysis up-to-now. 
It is shown that the RMC method 
allows to estimate from an observed spectrum 
the proper physical parameters of the absorbing gas and 
simultaneously an appropriate structure
of the velocity field parallel to the line-of-sight.

The application to the analysis of the H+D Ly$\alpha$ 
profile is demonstrated using
Burles \& Tytler [B\&T] data for QSO 1009+2956 where the
DI Ly$\alpha$ line is seen at $z_a = 2.504$.

The results obtained favor a {\it low} D/H ratio in this absorption
system, although our upper limit for the hydrogen isotopic ratio
of about $4.5\times10^{-5}$ is slightly higher than that of B\&T 
(D/H = $3.0^{+0.6}_{-0.5} \times 10^{-5}$).
We also show that the D/H and N(HI) values are, in general, 
{\it correlated}, i.e. the derived D-abundance may be
badly dependent on the assumed hydrogen column density. 
The corresponding confidence regions for an arbitrary and 
a fixed stochastic velocity field distribution are calculated.

\section{Introduction}

The measurement of deuterium abundance at high redshift from
absorption spectra of QSOs is the most sensitive test of physical
conditions in the early universe just after the era of Big Bang 
Nucleosynthesis [BBN]. The standard BBN model predicts  strong
dependence of the primordial ratio of deuterium to hydrogen nuclei
D/H on the cosmological baryon-to-photon ratio $\eta$
and the effective number of light neutrino species $N_\nu$
(e.g. Walker {\it et al.} 1991). The accuracy of the theoretically calculated
hydrogen isotopic ratio is rather high and for a given value of $\eta$,
D/H is determined with
a 15 \% precision (Sarkar 1996).
Therefore to be comparable with this small uncertainty in the predicted
D/H value astronomical measurements of the deuterium and hydrogen
column densities should be of a similar precision.

The analysis of interstellar (intergalactic) absorption lines is not,
however, an easy and unambiguous task, especially for the case
of optically thick lines. As demonstrated in a series of
papers (Levshakov \& Kegel 1997; Levshakov, Kegel \& Mazets 1997;
Levshakov, Kegel \& Takahara 1997; Papers I, II and III, hereinafter,
respectively) the main difficulty is connected with correlation
effects between different physical parameters that may occur in chaotic
media along the line-of-sight. For instance, if one considers the line 
formation process in the light of a point source, 
then the observed spectrum reflects
only one realization of the velocity field and, hence, large
deviations from the expected average intensity $\langle I_\lambda \rangle$
may occur if the correlation length of the velocity field $l$ is not
very small compared with the cloud size $L$. Paper I clearly
demonstrates that in general (i.e. when $l \neq 0$) 
the absorption line profile is asymmetric (skew)
and may look like a barely resolved blend whereas  
homogeneous density and temperature have been adopted in these calculations.

This fact becomes crucial for the analysis
of H+D Ly$\alpha$ spectra where hydrogen lines are 
always saturated. An example discussed in
Paper II suggests that the apparent scatter of the D/H ratio of more than a 
factor of 4 revealed in the ISM may be caused by an inadequate analysis.

The study of extragalactic hydrogen spectra  
may turn out to be more complicated.
Two limiting cases should be distinguished :

{\sl 1. Complete statistical ensemble}. If one observes  
UV-spectra of galaxies  when the spectrograph aperture covers
an essential part of the galactic surface, 
then $\langle I_\lambda \rangle$ should
reasonably correspond to the observations since this case is a good
approximation to the mathematical space averaging procedure. 
An example of the averaged 
mesoturbulent H+D Ly$\alpha$ spectra  
is considered by Levshakov \& Takahara (1996). 
It has been shown that the standard Voigt-fitting procedure 
applied to these spectra may yield either {\it higher}
or {\it lower} D/H value (up to a factor of 10) 
as compared with an adopted one. 

{\sl 2. Poor statistical ensemble}. If a hydrogen spectrum is observed 
in an intervening cloud along a QSO line-of-sight, and 
the quasar itself may be considered as a point-like source, 
then the light beam intercepts the absorbing region 
in {\it only one} direction. 
To obtain an unambiguous result in this case the structure 
of the stochastic velocity field must be known exactly.
The Voigt-fitting procedure may lead in this case to 
{\it underestimated} D/H values as will be shown below.

This report presents a self-consistent method which enables 
us to evaluate both the physical parameters of the gas cloud and 
the corresponding velocity field projection to the line-of-sight.

\section{The RMC Method and Results}

The main aim of the present study is the {\it inverse problem}, i.e. 
the problem to deduce physical parameters from an observed absorption spectrum
in the light of a point-like source.
For this we use the  same mesoturbulent model specified in
full detail in Paper II. The inverse problem is
always an optimization problem in which an objective function
is minimized. We use a $\chi^2$ method to estimate a goodness
of the minimization and a $\Delta(\chi^2)$ technique to draw
confidence regions.
Our model is fully defined by specifying the hydrogen column
density N(HI), the kinetic temperature T$_{kin}$, the ratio of the rms
turbulent velocity to the hydrogen thermal velocity
$\sigma_t/v_{th}$, and the $L/l$ ratio. In addition we have to
specify the distribution of the velocity component parallel
to the line of sight $v(s)$. Here $v(s)$ is
a continuous random function of the coordinate $s$. 
In the numerical procedure it is sampled at evenly spaced intervals
$\Delta s$. The number of intervals depends on the current values
of $\sigma_t/v_{th}$ and $L/l$, being typically $\sim 100$
(see Paper II, for details). Thus, to solve the inverse problem
we have to optimize the objective function in 
the parameter space of very large and variable (depending on
$\sigma_t/v_{th}$ and $L/l$) dimension.
The RMC method  based on the computational scheme invented
by Metropolis {\it et al.} (1953) proved to be adequate
in this case.

Contrary to the standard Monte Carlo procedure in which random
configurations of a given physical system are generated to
estimate its average characteristics, the RMC takes an
experimentally determined set of data and searches for
a random parameter configuration which reproduces the
observation. Applied to the analysis of absorption spectra,
the computational procedure is described in detail in Paper III.
Here we only note that the main idea was to divide the
parameter space into two parts : (1) the subspace of 
constant dimension which contains the physical parameters
like N(HI), D/H, {\it etc.}, and (2) the subspace of 
variable dimension containing the velocity components $v(s_i)$ .

We applied the procedure to the
H+D Ly$\alpha$ line observed by B\&T in the spectrum
of QSO 1009+2956 at $z_a = 2.504$. This spectrum was
selected since (i) it shows a well pronounced DI absorption,
and (ii) it was obtained with high spectral resolution and
signal-to-noise ratio. 

Using a {\it two}-cloud microturbulent model
B\&T derived the following parameters : total hydrogen
column density N(HI) = $2.9^{+0.4}_{-0.3} \times 10^{17}$
cm$^{-2}$, D/H = $3.0^{+0.6}_{-0.5} \times 10^{-5}$,
T$_{kin}$ = $2.1^{+0.1}_{-0.1} \times 10^4$ K \, and
$2.4^{+0.7}_{-0.7} \times 10^4$ K for the blue and red
subcomponents, respectively, the corresponding turbulent
velocities $3.2 \pm 0.4$ km s$^{-1}$ and
$2.3 \pm 1.4$ km s$^{-1}$, and a difference in the radial velocity
of 11 km s$^{-1}$.

\begin{figure*}
\vspace{0.0cm}
\hspace{-1.7cm}\psfig{figure=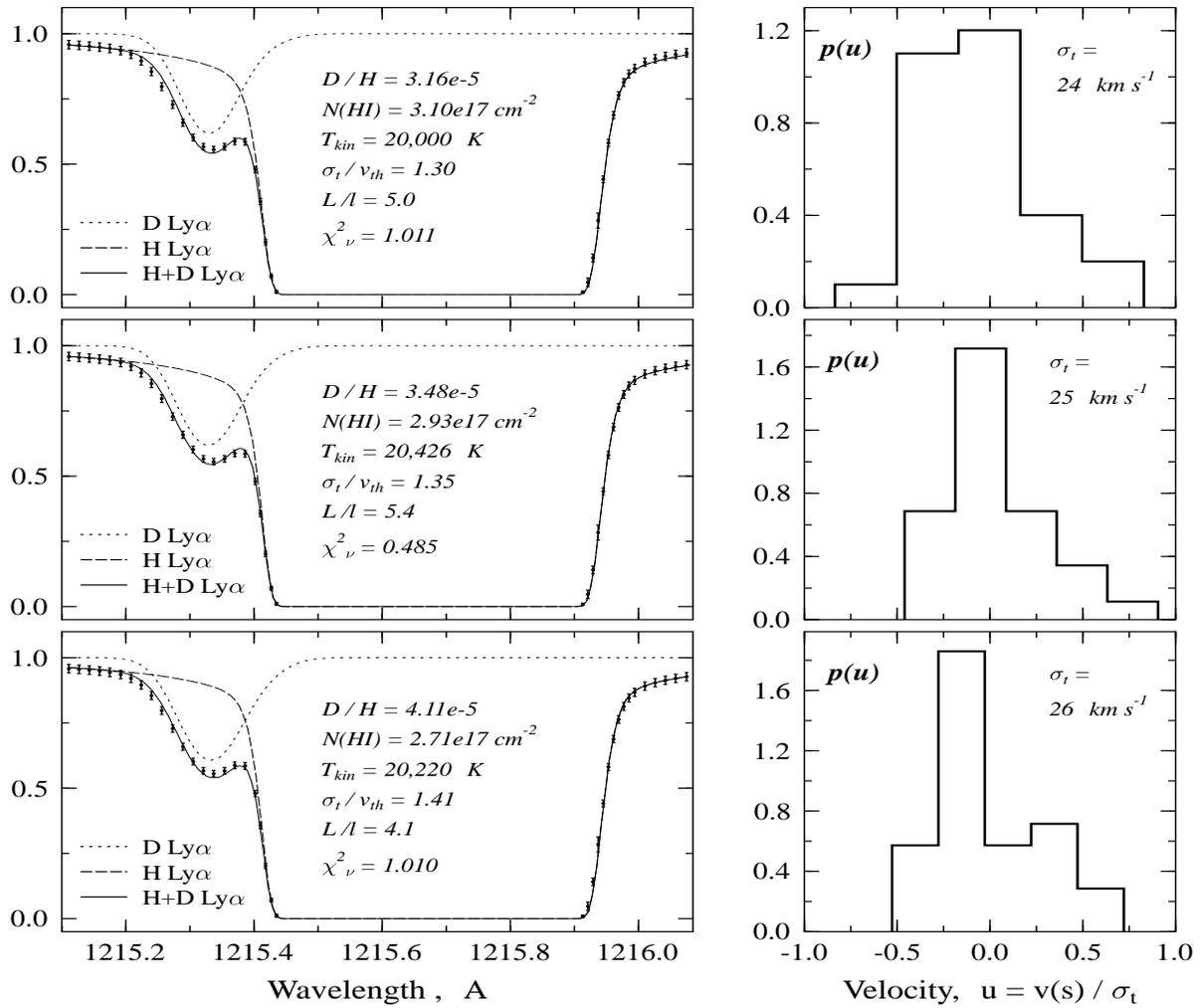,height=15.0cm,width=14.0cm}
\vspace{-4.0cm}
\caption[]{
A template H+D Ly$\alpha$ profile (dots) representing
the normalized intensities and their uncertainties in accord
with the B\&T data. The solid curves show 
the results of the RMC minimization.
The corresponding projected velocity field distributions
$p(u)$ are shown by histograms.}
\end{figure*}

\begin{figure*}
\vspace{2.0cm} 
\hspace{0.0cm}\psfig{figure=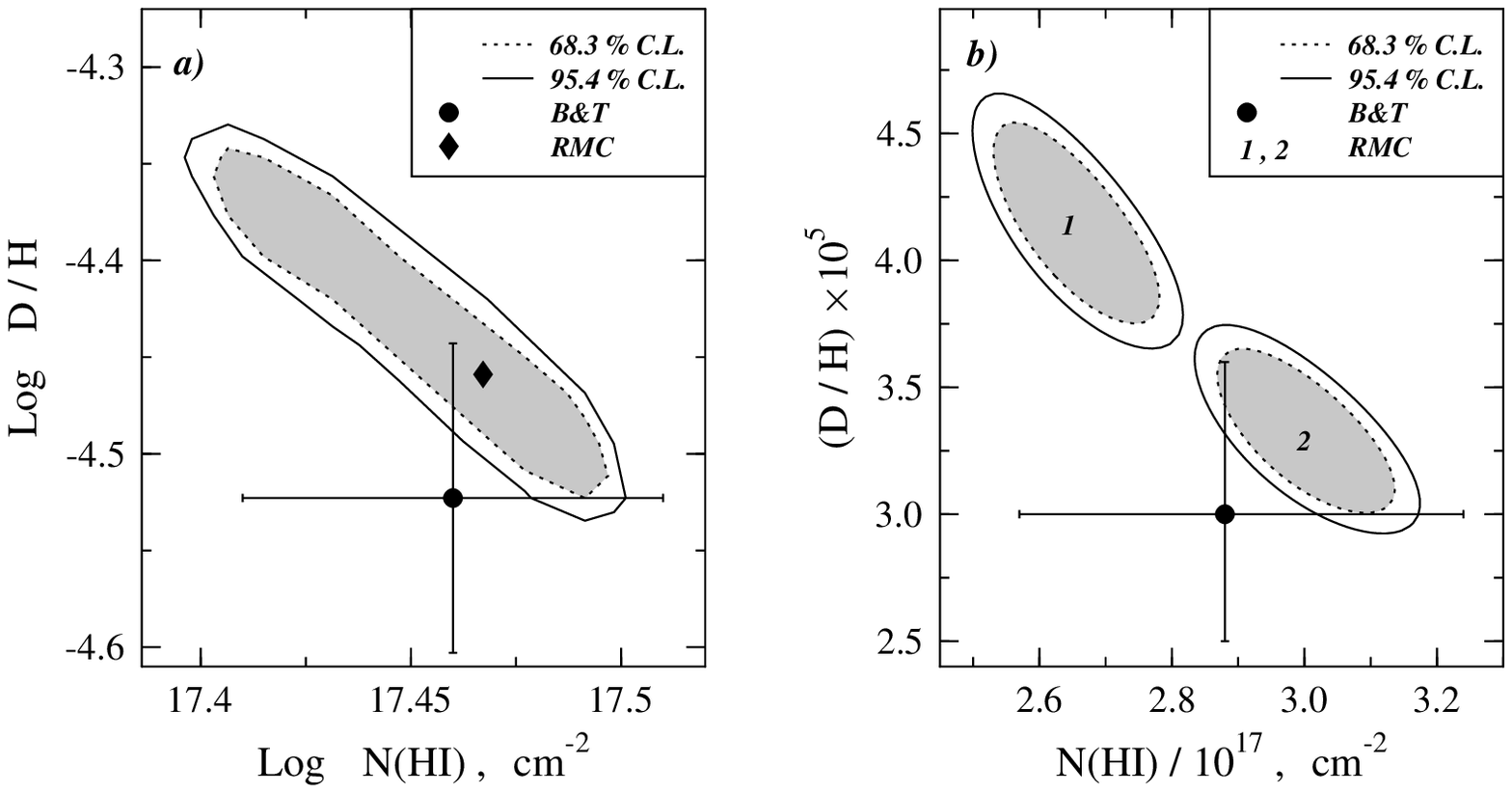,height=10.0cm,width=10.0cm}
\vspace{-5.2cm}
\caption[]{
({\bf a}) - Confidence regions for the plane 
``log N(HI) -- log D/H''
when the other parameters T$_{kin}$, $\sigma_t/v_{th}$
and $L/l$ are fixed (listed in the middle panel of Fig.~1)
but the configuration of the velocity field is free.
The B\&T and RMC best fitting parameters are labeled by
the filled circle and diamond, respectively. The B\&T error
bars correspond to $1\sigma$.
({\bf b}) - Confidence regions 1 and 2 for the fixed configurations
of the corresponding velocity fields.  
The fixed parameters T$_{kin}$, $\sigma_t/v_{th}$ and $L/l$ 
are listed in the lower and upper panels of Fig.~1,
respectively.}
\end{figure*}

We started with the calculation of a template
H+D Ly$\alpha$ spectrum based on the parameters listed
in Table 1 by B\&T. To simulate real data, we added
the experimental uncertainties to the template intensities
which were sampled in equidistant bins as shown in 
Fig.~1 by dots and corresponding error bars. We adopted
for the redshift the mean value of 2.504 and used a
{\it one}-component mesoturbulent model with a
homogeneous density and temperature.

Adequate profile fits for three different D/H values are
shown in Fig.~1 by solid curves. The estimated parameters
and corresponding values of $\chi^2$ (per degree
of freedom) are also listed for each solution.
The velocity distributions $p(u)$ leading to these profiles
[where $u = v(s)/\sigma_t$] are shown by histograms.

Although our model ({\it one} cloud with homogeneous
density and temperature accounting for a {\it finite}
correlation length in the stochastic velocity field)
is quite different from that of B\&T ({\it two} clouds,
stochastic velocities in each one considered in the
{\it microturbulent} approximation), we derive with
the RMC method in this particular case values for
N(HI), D/H, and T$_{kin}$ which are not significantly different
from those of B\&T. The rms turbulent velocity $\sigma_t$,
however, is substantially larger. For the additional
parameter $L/l$ we found a value of about 5, indicating
a substantial variation of the large-scale velocity field
along the line-of-sight. 

The best RMC solution with the smallest 
$\chi^2 \simeq 0.5$ (the middle panel in Fig.~1) is 
labeled by the filled diamond in Fig.~2a whereas the
filled circle is the B\&T result. In Fig.~2a, we show
confidence regions which are projected onto the plane 
``log N(HI) -- log D/H'' under the condition that the
{\it parameters} T$_{kin}$, $\sigma_t/v_{th}$ and $L/l$ are fixed, but
the velocity field {\it configuration} is free. The elongated
and declined shape of these confidence regions reflects
a correlation between D/H and N(HI), i.e.  
the scatter of the D/H values cannot be simply determined by
the projection of a given confidence region onto the
corresponding axis.

Another important result obtained is the existence of possible
multiple solutions with non-overlapping confidence regions.
This may be a real problem, if the multiplicity
scatters parameters significantly. To illustrate the problem
we draw two confidence regions labeled by numbers 1 and 2
(Fig.~2b).  These regions correspond to 
the RMC H+D profiles shown in the lower and upper panels
of Fig.~1 which present the limiting D/H values for the
mesoturbulent model. The regions 1 and 2 were calculated
in this case under the condition that 
T$_{kin}$, $\sigma_t/v_{th}$,  $L/l$ and the 
corresponding configurations of the velocity field are fixed.
Thus, if one succeeds to determine the velocity field
distribution $p(u)$ by including additional absorption lines
in the analysis, the accuracy of the D/H measurements 
may be improved significantly.

\section{Acknowledgments}
This work was supported by the Deutsche
Forschungsgemeinschaft and by the RFBR grant No.
96-02-16905-a.

\begin{center}
{\Large {\bf References}}
\end{center}

\bigskip\noindent
Burles, S. and Tytler, D. : 1996, 
`Cosmological deuterium abundance and the baryon density of\\
\hspace*{0.6cm}the Universe', astro-ph 9603070 [B\&T].\\
Levshakov, S.A. and Takahara, H. : 1996,
`The effect of spatial correlations in a chaotic velocity\\
\hspace*{0.6cm}field on the D/H measurements from QSO absorption spectra',
{\it Monthly Notices Roy.}\\
\hspace*{0.6cm}{\it Astron. Soc.} {\bf 279}, 651--660.\\
Levshakov, S.A. and Kegel, W.H. : 1997,
`New aspects of absorption line formation in\\
\hspace*{0.6cm}intervening turbulent clouds -- I. General principles',
{\it Monthly Notices Roy. Astron. Soc.}\\
\hspace*{0.6cm}{\bf 288}, 787--801, [Paper I].\\
Levshakov, S.A., Kegel, W.H. and Mazets I.E.: 1997,
`New aspects of absorption line formation\\
\hspace*{0.6cm}in intervening turbulent clouds -- II. Monte Carlo 
simulation of interstellar H+D Ly$\alpha$\\
\hspace*{0.6cm}absorption profiles', {\it Monthly Notices Roy. Astron. Soc.} 
{\bf 288}, 802--816, [Paper II].\\
Levshakov, S.A., Kegel, W.H. and Takahara F. : 1997,
`New aspects of absorption line formation\\
\hspace*{0.6cm}in intervening turbulent clouds -- III. The inverse problem in 
the study of H+D profiles',\\
\hspace*{0.6cm}{\it Monthly Notices Roy. Astron. Soc.} (submit.) [Paper III].\\
Metropolis, N. {\it et al.} : 1953,
`Equation of state calculations by fast computing machines',\\
\hspace*{0.6cm}{\it J. Chem. Phys.} {\bf 21}, 1087--1092.\\
Sarkar, S. : 1996, `Big bang nucleosynthesis and physics beyond the
standard model', {\it Rep. Prog.}\\
\hspace*{0.6cm}{\it Phys.} {\bf 59}, 1493--1609.\\ 
Walker, T.P. {\it et al.} : 1991, `Primordial nucleosynthesis redux',
{\it Astrophys. J.} {\bf 351}, 51--69.\\ 

\end{document}